\documentclass[sigconf, natbib=true]{acmart}

\AtBeginDocument{%
  }

\usepackage{arydshln}
\usepackage{lipsum}
\usepackage{tabularray}
\usepackage{multirow}
\usepackage{booktabs, caption, array}
\usepackage{algorithm}
\usepackage{algorithmic}
\usepackage{comment}
\usepackage{xcolor}
\usepackage{subcaption}
\usepackage{graphicx}
\usepackage{subcaption}
\usepackage{enumitem}
\usepackage{framed}
\usepackage{hyperref} 
\usepackage{amsmath}
\definecolor{shadecolor}{gray}{0.9}
\usepackage{comment}
\usepackage{xspace}
\usepackage{colortbl}
\usepackage{xcolor}
\newcommand{\paratitle}[1]{\vspace{1.5ex}\noindent\textbf{#1}}

\newcommand{\mymodel}{NHP-OAM\xspace}
\newcommand{\mydata}{OAMD\xspace}

\begin{document}

\title{To Search or to Recommend: Predicting Open-App Motivation with Neural Hawkes Process}

\author{Zhongxiang Sun}
\affiliation{
  \institution{Gaoling School of Artificial Intelligence\\Renmin University of China}
\city{Beijing}
  \country{China}
  }
\email{sunzhongxiang@ruc.edu.cn}

\author{Zihua Si}
\affiliation{%
  \institution{Gaoling School of Artificial
Intelligence\\Renmin University of China}
  \city{Beijing}\country{China}
  }
\email{zihua_si@ruc.edu.cn}

\author{Xiao Zhang}
\affiliation{
  \institution{Gaoling School of Artificial Intelligence\\Renmin University of China}
  \city{Beijing}
  \country{China}
  }
\email{zhangx89@ruc.edu.cn}

\author{Xiaoxue Zang}
\affiliation{%
  \institution{Kuaishou Technology Co., Ltd.}
  \city{Beijing}\country{China}
  }
\email{zangxiaoxue@kuaishou.com}

\author{Yang Song}
\affiliation{%
  \institution{Kuaishou Technology Co., Ltd.}
  \city{Beijing}\country{China}
  }
\email{yangsong@kuaishou.com}

\author{Hongteng Xu}
\author{Jun Xu}
\authornote{Jun Xu is the corresponding author.  Work partially done at Engineering Research Center of Next-Generation Intelligent Search and Recommendation, Ministry of Education.
\\
Work done when Zhongxiang Sun and Zihua Si were interns at Kuaishou.}
\affiliation{
  \institution{Gaoling School of Artificial Intelligence\\Renmin University of China}
    \city{Beijing}
  \country{China}
  }
\email{{hongtengxu, junxu}@ruc.edu.cn}

\renewcommand{\shortauthors}{Zhongxiang Sun et al.}

\begin{abstract}

Incorporating Search and Recommendation (S\&R) services within a singular application is prevalent in online platforms, leading to a new task termed \emph{open-app motivation prediction}, which aims to predict whether users initiate the application with the specific intent of information searching, or to explore recommended content for entertainment.
Studies have shown that predicting users' motivation to open an app can help to improve user engagement and enhance performance in various downstream tasks. However, accurately predicting open-app motivation is not trivial, as it is influenced by user-specific factors, search queries, clicked items, as well as their temporal occurrences. Furthermore, these activities occur sequentially and exhibit intricate temporal dependencies.
Inspired by the success of the Neural Hawkes Process (NHP) in modeling temporal dependencies in sequences, this paper proposes a novel neural Hawkes process model to capture the temporal dependencies between historical user browsing and querying actions. The model, referred to as \textbf{N}eural \textbf{H}awkes \textbf{P}rocess-based \textbf{O}pen-\textbf{A}pp \textbf{M}otivation prediction model (\mymodel), employs a hierarchical transformer and a novel intensity function to encode multiple factors, and open-app motivation prediction layer to integrate time and user-specific information for predicting users' open-app motivations.
To demonstrate the superiority of our \mymodel model and construct a benchmark for the Open-App Motivation Prediction task, we not only extend the public S\&R dataset ZhihuRec but also construct a new real-world \textbf{O}pen-\textbf{A}pp \textbf{M}otivation \textbf{D}ataset (\mydata). Experiments on these two datasets validate \mymodel's superiority over baseline models. Further downstream application experiments demonstrate \mymodel's effectiveness in predicting users' Open-App Motivation, highlighting the immense application value of \mymodel.
\end{abstract}

\ccsdesc[500]{Information systems~Users and interactive retrieval}

\keywords{Open-App Motivation, Neural Hawkes Process, Behavior Modeling}

\maketitle
\section{Introduction}

\begin{figure*}
    \centering
    \includegraphics[width=\textwidth]{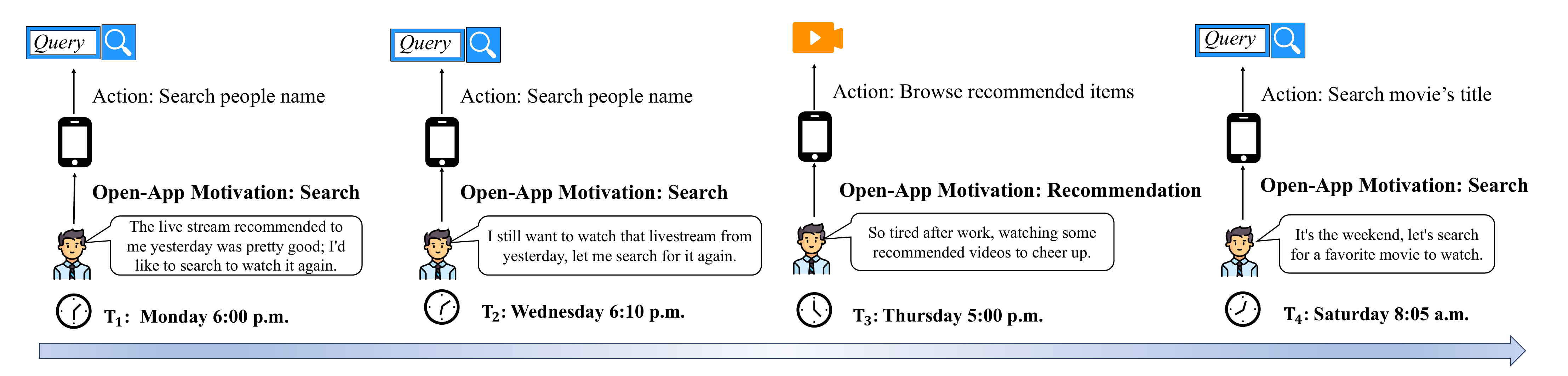}
    \caption{Examples of Open-App Motivation. Obtained through a structured questionnaire-based interview process with users.}
    \label{fig:Open-App_Motivation_Examples}
\end{figure*}

To combat information overload, recommender systems and search engines are integral in applications like online video and e-commerce platforms, allowing users to access information through active and passive modalities.
Consequently, this leads to the potential for users to open these applications with varying motivations.
\emph{Within these applications, two primary open-app motivations for users are to embrace suggestions from the recommendation system or employ the search engine for information seeking.}  Varied open-app motivations correspond to distinct needs and behavioral patterns.  As illustrated in Figure~\ref{fig:Open-App_Motivation_Examples}, a user might initiate the application driven by a specific intent, such as searching or perusing recommended videos. Notably, these motivations evolve, influenced by historical interactions as well as the user's lifestyle patterns.

An accurate prediction of the open-app motivation has significant meaning in the following aspects:
\begin{itemize}[leftmargin=*]
    \vspace{-0.1cm}
    \item \textbf{Enhancing user engagement:} Open-App motivation is vital for optimizing user engagement. In an experiment on a video platform with billions of users, altering the search modules from inconspicuous to a prominent position led to a significant rise in user interaction and elevated the related value of open-app to search from 0.1\% to 0.8\%~\footnote{An increase in the proportion of users open-app to search significantly improves the retention of low-activity users and can significantly increase the total usage time for users on the App.}. However, this change induced competition, with 87\% of the gains from the search and 13\% from a decrease in returns from recommended videos. Based on the prediction of the user's open-app motivation, we can dynamically adjust the search and recommendation modules of the Apps, thereby mitigating the competition between recommendation and search to improve user engagement.
    \item \textbf{Application in downstream tasks:} Open-App motivation plays a critical role in various downstream applications, including (1) Based on the prediction of open-app motivation, we can obtain more precise guidance of user interest sources (search or recommendation history) in recommended tasks that need to arouse user interest in a very short time, e.g., App open advertisement and message pop-up recommendation~\footnote{The introduction of App open advertisement and message pop-up recommendation:~\url{https://en.wikipedia.org/wiki/Pop-up_ad}.}~\cite{lan2023neon}. (2) Learn a better user representation, serving as a robust user characterization to enhance the recommendation system's effect.
\end{itemize}




This real-world problem relates to two research topics: User Retention Prediction~\cite{he2023concept,li2022automatically,zhang2023efficiently,chang2023latent,fan2019metapath} and User Intent Prediction~\cite{xuhybrid,lieffective,cai2023reinforcing,kim2018utilizing,kim2020revisit}. User Retention Prediction forecasts if a returning user will engage with an App again. The prediction does not consider the fine-grained specifics of the open-app motivation, i.e., whether for searching or embracing recommended information. User Intent Prediction predicts what intents a user might have while utilizing an App, representing a higher-level understanding of user intents within the App's environment. Intent types are varied and hard to define explicitly. Different Apps may focus on different intents, such as purchasing in e-commerce Apps or entertainment in video Apps. We study why users open an App and acknowledge they may search or embrace recommendations for a particular intent. This leads to a new, widely applicable problem across various Apps.

\noindent \textbf{Challenges in Open-App Motivation Prediction.}
Open-App motivation prediction faces significant challenges, which we further substantiate with behavioral patterns observed in real-world users' data (\textsection~\ref{sec:data_analysis}):
\begin{itemize}[leftmargin=*]
    \vspace{-0.1cm}
    \item \textbf{Time-User bias:} Open-App motivation varies based on time and individual users. Our data analysis reveals the ratio of search to recommendations exhibits an apparent \textbf{Periodicity} within each 24-hour day, as well as between workdays and rest days of the week. Combined with the case study about individual users shown in Figure~\ref{fig:Open-App_Motivation_Examples}, this validates the challenge of time-user bias.
    
    \item \textbf{Multiple factors:} The motivation is influenced by various factors, e.g., search queries and clicked recommended items in history logs. Our findings (\textsection~\ref{sec:data_analysis}) indicate two key patterns under this challenge. First, a phenomenon of \textbf{Repeat-Query} behavior suggests that users may open the App multiple times to search for the same query. Second, we observe a positive \textbf{Relevance} between the ratio of searches to clicked recommended items in a user's previous session and the motivation to search when opening the App the next time, further highlighting the multifaceted influences affecting Open-App motivation.

\end{itemize}

To address the challenges mentioned above, combining our analysis of users' open-app motivation's behavioral patterns, we propose a novel Neural Hawkes Process~\cite{zhang2020self,zuo2020transformer,mei2017neural} (NHP)-based model, called NHP-OAM, to predict users' open-app motivations~(\textsection~\ref{sec:model} ). Considering that open-app motivation exhibits Periodicity and Repeat-Query features and the fact that the time intervals at which users open the App are not fixed, the neural Hawkes process can effectively capture the clear periodicity and repeated query features found in past open-app motivations without needing to assume consistent sampling time intervals like in time series modeling~\cite{lim2021time}. Unlike previous NHP models, considering that open-app motivation is influenced by multiple factors, we utilize a hierarchical transformer model as the history encoder which can not only learn the past open motivation sequence but also learn the session-level and history-level contextual representations of S\&R behaviors. In the design of the intensity function for \mymodel, to capture the Relevance feature, we propose an intensity function that is aware of both the S\&R behaviors' ratio and representations learned from the hierarchical transformer model. Finally, considering the temporal factor and users' different habits that significantly impact open-app motivation, we employ a time gate to fuse temporal information with the hierarchical user history representation and specific user embedding to predict users' open-app motivation.

The contributions of this paper are summarized as follows:

\begin{itemize}[leftmargin=*]
    \vspace{-0.1cm}
    \item We pioneer the study of open-app motivation prediction, which is a critical problem in real-world Apps that integrate S\&R services but has not been well explored.

    \item We propose the \mymodel that effectively leverages the behavioral patterns underlying open-app motivation to address associated challenges. Utilizing the properties of NHP, we capture Periodicity and Repeat-Query features, aiding in resolving the Time-User bias challenge. A hierarchical transformer model is employed to capture the influence of Multiple factors on open-app motivation. Additionally, we introduce an intensity function aware of the Relevance features.

    \item We construct and open-source the first real-world \textbf{O}pen-\textbf{A}pp \textbf{M}otivation \textbf{D}ataset (\mydata), which contains users' S\&R behaviors, as well as the motivation for each session when opening the App. To further validate the effectiveness of our model, we also extend the public S\&R datasets-ZhihuRec~\footnote{We will release these datasets after acceptance.}. Experiments on these two datasets validate \mymodel's superiority over baseline models. Further downstream  Application experiments demonstrate \mymodel's effectiveness in predicting users' open-app motivation, underlining its application value.

\end{itemize}



\section{Related Work}
\paratitle{User Retention Prediction.}
User Retention Prediction aims to forecast the likelihood of a user returning to an app for further interactions.  Various techniques have been developed to tackle this issue. \citet{xuhybrid} utilized a hybrid approach for user retention in a short video platform. \citet{lieffective} aligns with \cite{xuhybrid} in their use of ensemble methods, but specifically focused on multichannel time series for a long video Internet company. Diverging from ensemble methods, \citet{cai2023reinforcing} leveraged reinforcement learning for long-term user retention in Kuaishou~\cite{cai2023reinforcing}. In contrast, \citet{kim2020revisit} as well as \citet{kim2018utilizing} extended the focus to offline retail, using deep survival analysis and Wi-Fi fingerprinting, respectively.
In contrast to User Retention Prediction, our work on Open-App Motivation Prediction aims to identify the specific motivation why a user opens an app, adding a fine-grained study compared to existing studies.

\paratitle{User Intent Prediction.}
User Intent Prediction has been extensively studied in various domains. \citet{he2023concept} focused on using a concept knowledge graph to characterize user intent at Alipay explicitly. In contrast, \citet{li2022automatically} designed AutoIntent to discover implicit consumption intents from user data at Meituan. \citet{zhang2023efficiently} introduced Atten-Mixer, which leverages multi-level user intent for session-based recommendation. \citet{chang2023latent} proposed a probabilistic approach using variational autoencoders to infer latent user intent in sequential recommenders.  \citet{fan2019metapath} utilized a metapath-guided graph neural network to recommend user intent in mobile e-commerce platforms like Taobao. Our work shifts focus from traditional User Intent Prediction to Open-App Motivation Prediction. Instead of forecasting multiple in-app intents, we aim to pinpoint why a user initially opens an app for search or recommendation. This provides a fine-grained insight into user behavior, which is essential for integrating S\&R apps.


\section{\mbox{Empirical study}}
In this section, we first give the task definition of open-app motivation prediction. Then we make several observations about open-app motivation on the real-world dataset (\mydata), which serves as the foundation of our model.

\subsection{Task Definition}
Let $\mathcal{U}$ be a set of users and $\mathcal{I}$ be a set of items. We consider user queries by introducing an additional query set $Q$ and word vocabulary $\mathcal{W}$, where a query $q \in Q$ consists of a list of words $\left[w_1, \ldots, w_{|q|}\right], w \in \mathcal{W}$.
The interactions of user $u \in \mathcal{U}$ sorted chronologically is organized as a sequence of $N$ sessions~\footnote{A session is defined as all the user's interactions from when they open the App to when they close it.} $\mathcal{S}  = \left\langle \left\{ \mathcal{S}_1 , m_1 , t_1  \right\}, \left\{ \mathcal{S}_2 , m_2 , t_2  \right\}, \cdots, \left\{ \mathcal{S}_N , m_N , t_N  \right\} \right\rangle$, where $t_n $  is the time when the user $u$ opens the App at session $\mathcal{S}_n  $, and $m_n $ is the open-app motivation for session $\mathcal{S}_n $, where 
\begin{equation}
m_n   = \begin{cases}1, & \text{open-app motivation is search} \\ 0, & \text{open-app motivation is recommendation} \end{cases} .
\end{equation}
Each session $\mathcal{S}_n =\left\{s_{n, 1} , s_{n, 2} , \cdots, s_{n,\left|\mathcal{S}_n \right|} \right\}$, where $\left|\mathcal{S}_n \right|$ is the length of this query-aware heterogeneous sequence. $s_i  $ can be an item interaction or a query action. Let $\delta$ indicates whether $s_i  $ at $i$-th step is a query or an item interaction:
\begin{equation}
s_i   \in \begin{cases}\mathcal{I}, & \text { if } \delta\left(s_i  \right)=0 \\ Q, & \text { otherwise. }\end{cases}  
\end{equation}
Let $t(\mathcal{S}_{n}) = \left\{t_{n, 1} , t_{n, 2} , \cdots, t_{n,\left|\mathcal{S}_n \right|} \right\} $ represent the times of interactions or queries in session $\mathcal{S}_{n}$.

\noindent \textbf{Goal.} Given a user $u \in \mathcal{U}$ and her/his historical sessions with open-app motivations  $\mathcal{S} $, we want to predict the open-app motivation $m_{N+1}$ that $u$ will most probably have in the next session $\mathcal{S}_{N+1} $ at time $t_{N+1}  $.

\subsection{Dataset Description}
\label{sec:oamdata describe}

Open-App motivation prediction is a novel task, and there isn't a publicly available dataset for it. To bridge this gap, we collected search and recommendation behaviors as well as the motivation for each session when users open the App on a real video platform used by over one billion users~\footnote{We select users from who searched in the month prior to our data collection.}, called \mydata. Search and recommendation behaviors are directly obtained from the user behavior logs. As for the open-app motivation label, we obtain it from the internal data center of the video platform. \emph{This label is defined by whether the user actively searches within 30 seconds of opening the App. If true, the open-app motivation is ``search''; otherwise, it is ``recommendation''. }This classification method is rooted in the platform's comprehensive data analytics and business strategy.  Considering commercial confidentiality and privacy issues, we are unable to display the specific analysis data and content. 
To further elucidate the rationale behind this label, let's delve into two distinct perspectives:
\textbf{Qualitative Perspective:} Typically, users with a deliberate intent to search will begin their search soon after App activation. Given potential latency due to factors like network or device performance, a window of 30 seconds has been determined as a reasonable threshold to gauge users' motivation.
\textbf{Quantitative Analysis:} Data analysis in the dataset shows that in sessions where users actively searched within the first 30 seconds, the ratio of searched clicked items to recommended clicked items stands at \textit{0.4082}. This is notably higher than the \textit{0.0474} ratio in sessions without an active search in that window. This stark difference underscores the reliability of our 30-second criterion as an indicator of a user's open-app motivation. 

More detailed statistics of \mydata are summarized in \textsection~\ref{sec:dataset}.

\begin{figure}
    \centering
    \begin{subfigure}{0.98\linewidth}
    \centering
    \includegraphics[width=\textwidth]{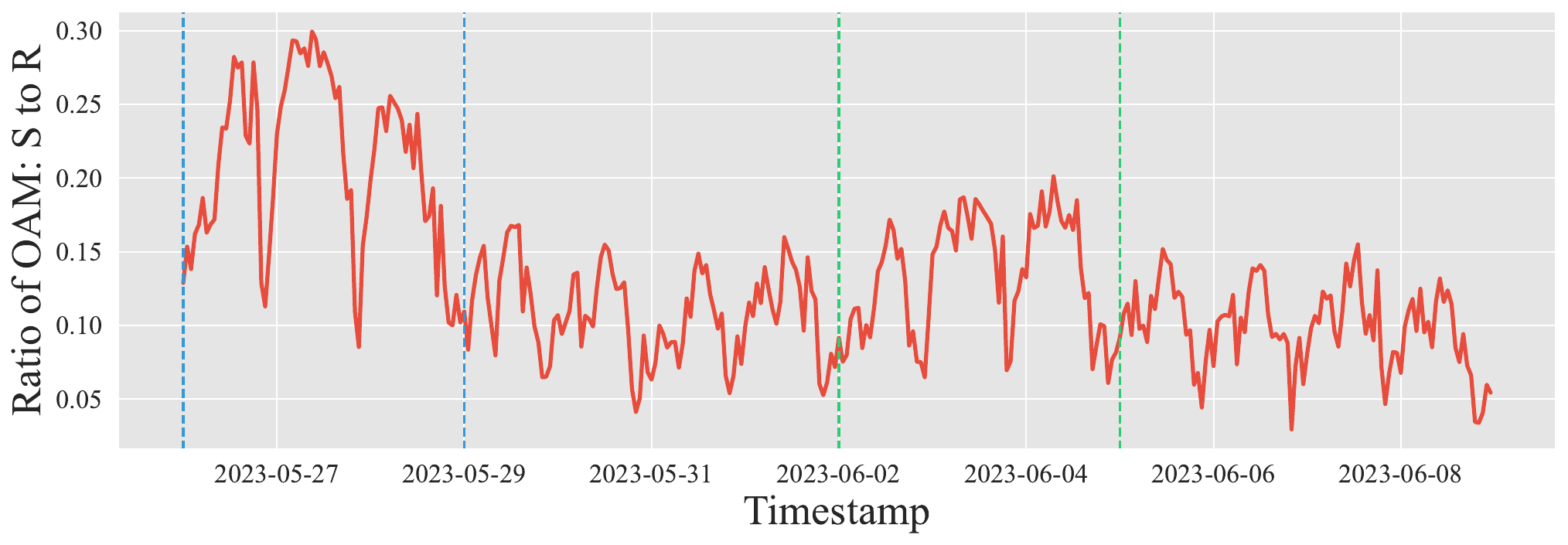}
    \subcaption{Overall Trend:  areas between the two dashed lines are weekends.}
    \label{fig:all_hour_trend}
    \end{subfigure}
    \centering
    \begin{subfigure}{0.49\linewidth}
        \centering
        \includegraphics[width=\textwidth]{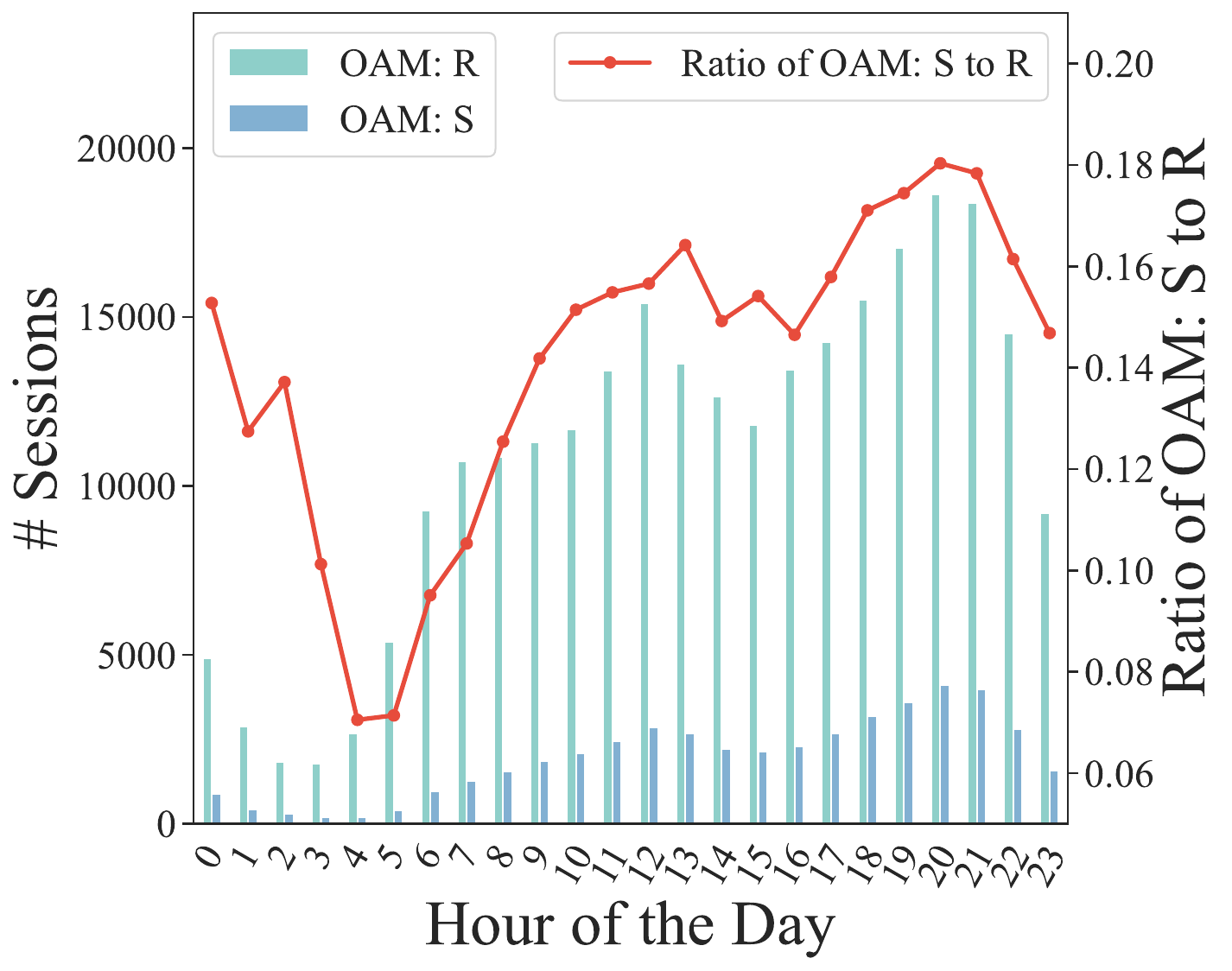}
        \subcaption{Hourly Trend.}
        \label{fig:hour_trend}
    \end{subfigure}
    \begin{subfigure}{0.49\linewidth}
        \centering
        \includegraphics[width=\textwidth]{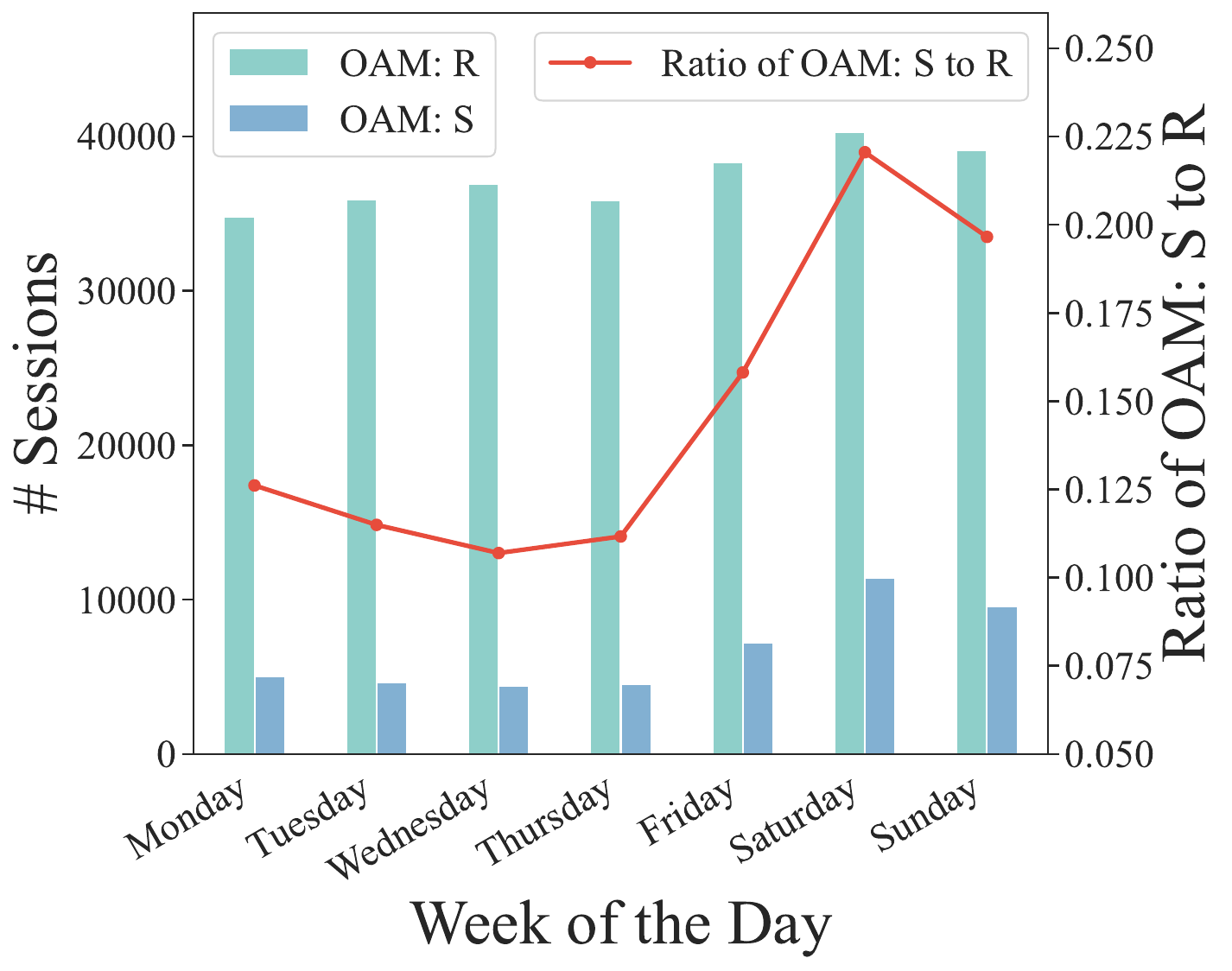}
        \subcaption{Weekly Trend.}
        \label{fig:week_trend}
    \end{subfigure}
    \caption{Periodicity statistics in Open-App Motivation. OAM: S denotes the Open-App Motivation: Search. OAM: R denotes the Open-App Motivation: Recommendation.}
\label{fig:statistics}
\end{figure}

\subsection{Behavioral Patterns Behind Users' Open-App Motivation}
\label{sec:data_analysis}
In this section, we study the behavioral patterns behind users' open-app motivation in the real-world dataset~\mydata. We aim to gain insights that show what a model should focus on when predicting users' next open-app motivation.

\subsubsection{Periodicity in Open-App Motivation}

We study how the ratio of search to recommendation in open-app motivations varies over time. Specifically, we analyze the hourly and daily trends in the ratio of searches to recommendations. Figure~\ref{fig:all_hour_trend} shows that users' open-app motivations have daily and weekly patterns in the search-to-recommendation ratio. Figure~\ref{fig:week_trend} reveals higher open-app to search on weekends (Friday nights,
Saturdays, and Sundays) than on weekdays. Figure~\ref{fig:hour_trend} shows users prefer to open apps for searches during midday and evenings. The trends for recommendations in users' open-app motivations are the exact opposite of these patterns. This demonstrates the importance of considering the temporal dependencies in open-app motivation, which needs to be emphasized in our model predicting users' open-app motivation.

\begin{figure}
    \centering
    
    \begin{subfigure}{0.49\linewidth}
        \centering
        \includegraphics[width=\textwidth]{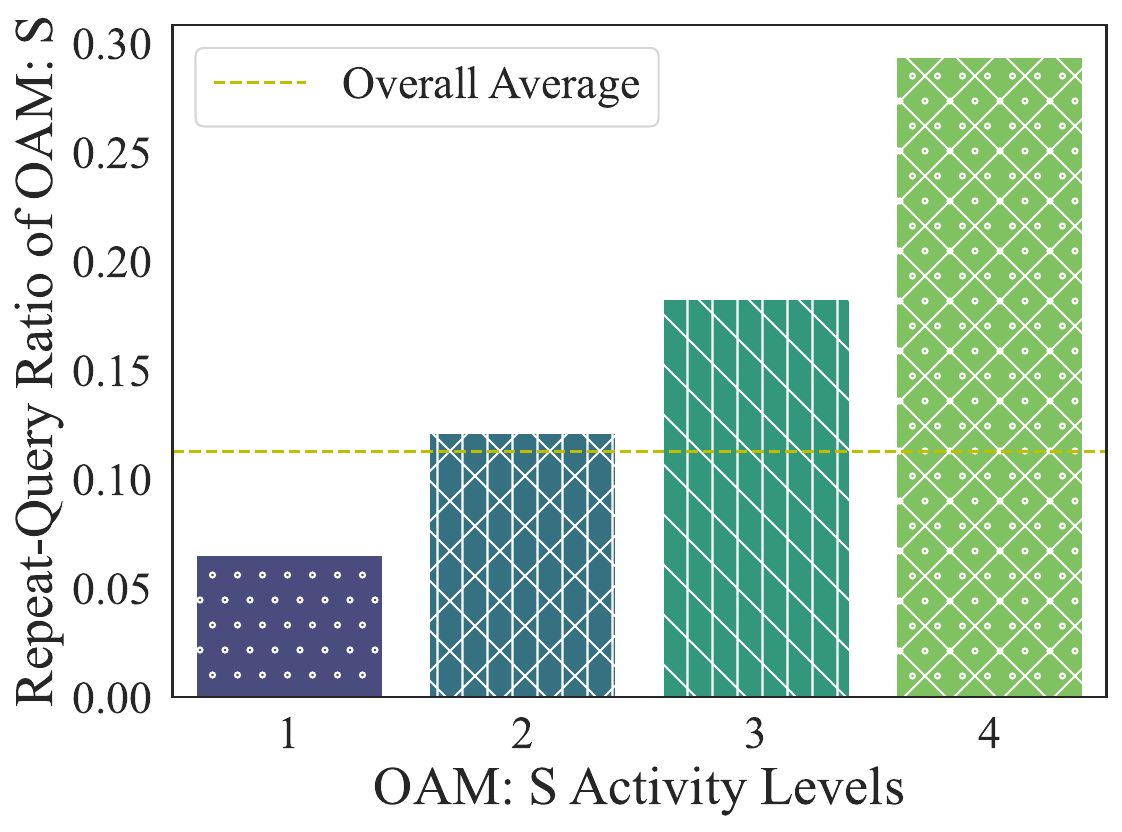}
        \caption{Repeat-Query}
        \label{subfig:repeat_query}
    \end{subfigure}
    \hfill
    \begin{subfigure}{0.49\linewidth}
        \centering
        \includegraphics[width=\textwidth]{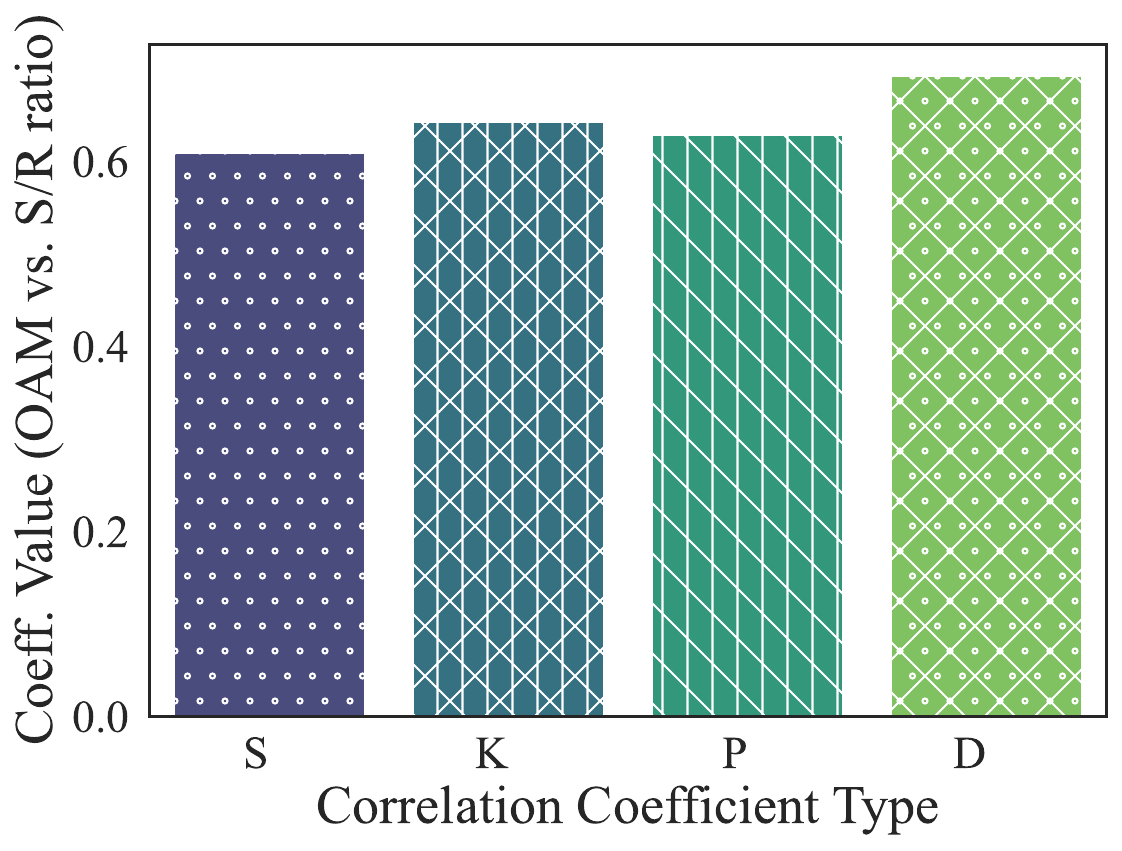}
        \caption{Relevance}
        \label{subfig:coeff_value}
    \end{subfigure}
    
    \caption{Open-App Motivation Features. (a) Repeat-query statistics in open-app motivation to search. Users are divided into four activity groups by Open-App Motivate: Search (OAM: S) activity levels (high values refer to high activity levels). The overall average denotes the repeat-query ratio in all the users.   (b) Relevance statistics between the ratio of clicked search list to clicked recommended list in users' past sessions and their motivation to search when they open the APP next time.}
    \label{fig:statistics_2}
\end{figure}

\subsubsection{Repeat-Query in Open-App Motivation} The Repeat-Query in the open-app motivation task is defined as the phenomenon where users open the app due to the same search query. We first categorize users into four bins of equal quantity based on their ``open-app to search'' activity level. Then, within different user bins, we calculated the average proportion for each user of all their ``open-app motivations to search'' due to searching the same query. As shown in Figure~\ref{subfig:repeat_query}, the higher the activity level, the higher the repeat-query ratio, and the overall average repeat-query ratio per user is also a comparatively high proportion. This finding highlights the need to use models that can understand a user's past actions to better predict their open-app motivations.

\subsubsection{Relevance in Open-App Motivation} Considering the potential relevance between the user's behavior within a session and the motivation to open the app, we investigate how the ratio of searches to clicks on recommended items influences a user's motivation to next open the app. We employ four correlation coefficients---Spearman (S), Kendall (K), Pearson (P), and Distance (D)---for a comprehensive analysis. Distance Correlation ranges within $[0, 1]$, while the others are in $[-1, 1]$. A higher value signifies a stronger correlation for all metrics. Figure~\ref{subfig:coeff_value} reveals a positive correlation across all metrics between past search-to-click ratios and future open-app motivation. Thus, the relevance feature can serve as a direct feature for determining a user's next open-app motivation.

\section{Our Approach: \mymodel}
\label{sec:model}
Based on the empirical observations above, we propose a Deep
Hawkes Process-based Open-App Motivation (\mymodel) prediction model. Some preliminaries about the Neural Hawkes Process are introduced first. Then we detailedly describe model definition and parameter learning.

\begin{figure*}
    \centering
    \includegraphics[width=0.98\linewidth]{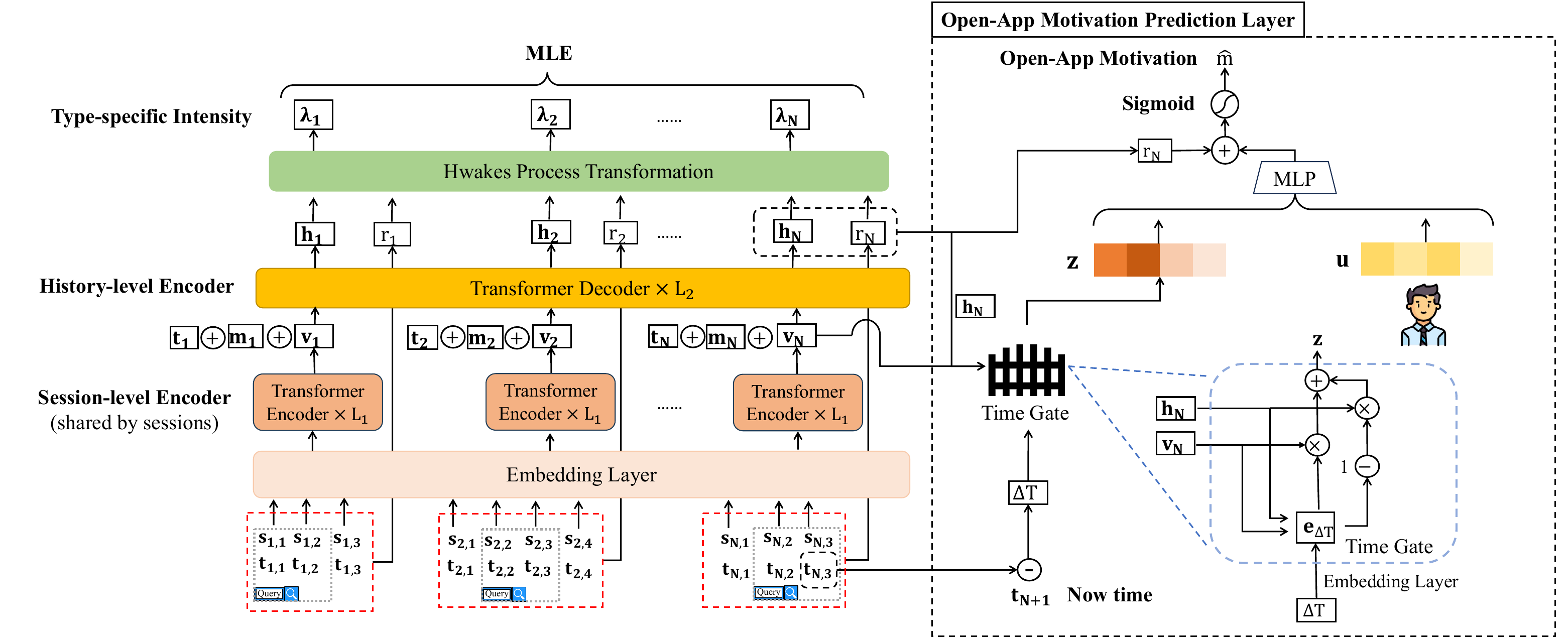}
    \caption{
    The overall architecture of the proposed model \mymodel, featuring Session-level Encoder and History-level Encoder in~\textsection~\ref{sec:History_Encoder}, Type-specific Intensity Function  in~\textsection~\ref{sec:Intensity_Function} and Open-App Motivation Prediction Layer in~\textsection~\ref{sec:Open-App Motivation Prediction Layer}.}
    \label{fig:model_graph}
\end{figure*}

\subsection{Preliminaries about Neural Hawkes Process}
Formally, a \textit{Hawkes Process}~\cite{hawkes1971spectra} is a temporal point process where one event can boost the likelihood of future events. Extending this, the \textit{Neural Hawkes Process} incorporates neural networks, offering a more flexible way to model temporal dependencies of events, thus capturing more complex patterns in data. The realization of a neural Hawkes process consists of a list of discrete events localized in time, which is denoted as $\mathcal{S} $ in the context of open-app motivation. 


Given the history time of past events \( \mathcal{S}   \), the neural Hawkes process begins by employing a history encoder—such as a Transformer~\cite{zhang2020self, zuo2020transformer} or LSTM~\cite{mei2017neural}—to encode these past events into a history representation. This representation is then used in a  conditional intensity function  (CIF) $ \lambda\left(t \mid \mathcal{T}_t\right) $ that provides a stochastic model for the time of the next event considering all times of previous events, where \begin{equation}
    \mathcal{T}_t=\left\langle \left\{\mathcal{S}_n, m_n, t_n  \right\}: t_n<t \right\rangle
\end{equation} is the history up to time $t$. The CIF captures the impact of previous events and allows for intricate modeling of dependencies through the history representation. Previous works have modeled the CIF using an exp-decay function with softplus~\cite{mei2017neural, zhang2020self} or a linear function with softplus~\cite{zuo2020transformer}.
In the context of open-app motivation, the probability for the occurrence of a new open-app motivation given the history time \( \mathcal{T}_t \) within a small time window \( [t, t+ dt) \) is given by the conditional intensity function:
\begin{equation*}
\lambda_{m}\left(t \mid \mathcal{T}_t\right)  \, dt = \mathbb{P}\left\{\text{event $m$ in } [t, t+ dt) \mid \mathcal{T}_t\right\}.
\end{equation*}

The conditional intensity function for the entire open-app motivation sequence is defined by:
\begin{equation*}
    \lambda\left(t \mid \mathcal{T}_t\right)=\sum_{m=0}^1 \lambda_m\left(t \mid \mathcal{T}_t\right).
\end{equation*}

Given the CIF, the probability density function (PDF) of the type-$m$ open-app motivation is:
\begin{equation}
    p_{m}\left(t \mid \mathcal{T}_t\right)=\lambda_{m}\left(t \mid \mathcal{T}_t\right) \exp \left(-\int_{t_j}^{t} \lambda_{m}\left(\tau \mid \mathcal{T}_\tau\right) d \tau\right).
\end{equation}


\subsection{\mymodel: History Encoder}
\label{sec:History_Encoder}
Previous neural Hawkes models used historical events and time data to predict future events~\cite{zhang2020self,mei2017neural,zuo2020transformer}. However, in our task, the motivation to open an App is influenced by both the queries searched and the items clicked on in the historical logs. To capture the rich contextual information within a user's session, as shown in~\autoref{fig:model_graph}, we designed a hierarchical transformer model as \mymodel's history encoder. First, using the transformer encoder module, the user's query and click interaction behaviors within the session are encoded into a session-level representation. Subsequently, the past session's session-level information is combined with the motivation to open the App and the time information. This is fed into the transformer decoder module of the hierarchical transformer model, ensuring that future data is not leaked, resulting in history-level representations that can be used to compute the intensity function. Next, we'll detail the hierarchical transformer model.


\subsubsection{Embedding Layer.}
\label{sec:embedding_layer}
We use embeddings $\mathrm{M_{U}} \in \mathbb{R}^{ |\mathcal{U}| \times d}, \mathrm{M}_{I} \in \mathbb{R}^{|\mathcal{I}| \times d}, \mathrm{M}_{W} \in \mathbb{R}^{|W| \times d} , \mathrm{M}_{M} \in \mathbb{R}^{2 \times d}, \mathrm{B} \in \mathbb{R}^{2 \times d}$   to represent $d$-sized users, items, query words, open-app motivation type, interaction type, respectively. We represent the inputs: (1) User: Given user $u$, we look up corresponding embeddings $\mathrm{M}_{U}$.  (2) Item: Given item $i$, we look up corresponding embeddings $\mathrm{M}_{I}$. (3) Open-App motivation: Given item $m$, we  look up corresponding embeddings $\mathrm{M}_{M}$. (4) Query: For query $q=\left[w_1, \ldots, w_{|q|}\right]$, we retrieve corresponding word embeddings and adopt an average pooling operation to get our query representation. (5) Interaction type: We look up $B \in \mathbb{R}^{2 \times d}$ to get embeddings for different interaction types (i.e., item or query). (6) Time: For the timestamp $t$, we apply a positional encoding method following~\cite{zuo2020transformer}: 
\begin{equation}
    \left[e_t\right]_i= \begin{cases}\cos \left(t / 10000^{\frac{i-1}{d}}\right) & \text { if } i \text { is odd } \\ \sin \left(t / 10000^{\frac{i-1}{d}}\right) & \text { if } i \text { is even }\end{cases},
\end{equation}
where $e_{t}$ denotes the temporal embedding and $i$ is the step of the interaction corresponding to $t$.

\subsubsection{Session-level Encoder.}  Given a user $u \in \mathcal{U}$ and one of her/his heterogeneous historical session $\mathcal{S}_n  $, the task of the session-level encoder is to generate the session embeddings $\textbf{v}_n $ representing $u$'s intention in the current session, using transformer encoder layers. Next, we will delve into the computation of \( \textbf{v}_n  \). 

For the items or queries $\left\{s_{n, 1}, s_{n, 2}, \cdots, s_{n, |\mathcal{S}_{n}|}\right\}$ of session $\mathcal{S}_n$, we assemble their embeddings as follows: if $\delta(s_{n, i})$ equals to 0, we use \hyperref[sec:embedding_layer]{\textit{Embedding Layer (2)}}; otherwise,  we use \hyperref[sec:embedding_layer]{\textit{Embedding Layer (4)}}. The assembled embeddings form an embedding matrix $\mathbf{E}_{n}^{s}=\left[\mathbf{e}_{n, 1}, \mathbf{e}_{n, 2}, \cdots, \mathbf{e}_{n, |\mathcal{S}_n|}\right]^{\mathbf{\top}} \in \mathbb{R}^{ |\mathcal{S}_n|\times d}$. Similarly, we get the time embeddings $\mathbf{E}_{n}^{t}$ according to \hyperref[sec:embedding_layer]{\textit{Embedding Layer (6)}} and interaction type embeddings $\mathbf{E}_{n}^{b}$ according to \hyperref[sec:embedding_layer]{\textit{Embedding Layer (5)}}. Combining these three embeddings, we get the session embedding matrix:
\begin{equation*}
\mathbf{E}_{n}=\mathbf{E}_{n}^{s}+\mathbf{E}_{n}^{t}+\mathbf{E}_{n}^{b},
\end{equation*}
where $\mathbf{E}_{n} \in \mathbb{R}^{ |\mathcal{S}_n|\times d}$ and $+$ denotes element-wise addition.
Then, We build $L_{1}$ Transformer Encoder~\cite{vaswani2017attention} blocks as the Session-level Encoder to learn the session-level representation $\textbf{v}_{n}$:
\begin{equation}
\textbf{v}_{n}   = \text{MEAN}\left(\hat{\mathbf{E}}_{n}\right);
\hat{\mathbf{E}}_{n} = \text{Encoder}(\mathbf{E}_{n}),
\end{equation}
where $\textbf{v}_{n}   \in \mathbb{R}^{d}$, $\hat{\mathbf{E}}_{n}\in \mathbb{R}^{ |\mathcal{S}_n|\times d}$ and MEAN$\left(\right)$ denotes mean pooling. The last session embedding, \( \textbf{v}_{N}   \), is utilized as the present short-term intent embedding, as it reflects a user's immediate preference.

\subsubsection{History-level Encoder.}
 Given a user $u \in \mathcal{U}$ and all her/his heterogeneous historical sessions $\mathcal{S}  $, the task of the history-level encoder is to generate the history representation $\mathbf{H}  $ for the intensity function and long-term embedding for the open-app motivation prediction by using transformer decoder layers. Using the \textit{Session-level Encoder}, \hyperref[sec:embedding_layer]{\textit{Embedding Layer (3)}} and \hyperref[sec:embedding_layer]{\textit{Embedding Layer (6)}} to encode $\mathcal{S}_n  $, $m_n $,  $t_n $, respectively, we can get the embedding of  $\mathcal{S}  $:
 \begin{equation*}
     \mathbf{E}_{\mathcal{S}  }= \left [ \textbf{v}_1 +\textbf{e}_{m, 1}  + \textbf{e}_{t, 1} ,  \textbf{v}_2  + \textbf{e}_{m, 2}  + \textbf{e}_{t, 2} , \ldots,  \textbf{v}_N  + \textbf{e}_{m, N}  + \textbf{e}_{t, N}  \right ]^{\mathbf{\top}},
 \end{equation*}
 where $\mathbf{E}_{\mathcal{S}  } \in \mathbb{R}^{N \times d}$ is the embedding matrix of all the user's past open-app motivation sequence.
 The embedding matrix $\mathbf{E}_{\mathcal{S}  }$ is then fed through $L_2$ Transformer Decoder~\cite{vaswani2017attention} blocks as the History-level Encoder, generating time-aware hidden representations $\mathbf{H}$ of the input open-app motivation sequence:
 \begin{equation}
     \mathbf{H} = [\mathbf{h}(t_{1}), \mathbf{h}(t_{2}), \cdots, \textbf{h}(t_{N})]^{\mathbf{\top}} = \text{Decoder}(\mathbf{E}_{\mathcal{S}  }),
 \end{equation}
where $\mathbf{H} \in \mathbb{R}^{N \times d}$ is composed of the hidden representations of the input sequence and \( d \) is the dimension of each representation. Each \( \mathbf{h}(t_{n}  ) \) corresponds to the hidden representation of the \( n \)-th input at time $t_{n}$ in the open-app motivation sequence for user \( u \), which contains all historical information from the beginning to the \( n \)-th input. The matrix \( \mathbf{H} \) is utilized to compute the intensity function. Its last position, \( \mathbf{h}(t_{N}) \), includes all historical actions of the current user and can be used to represent the long-term user intent.

\subsection{\mymodel: Intensity Function}
\label{sec:Intensity_Function}
Since the intensity function of Hawkes processes is history-dependent, we use \(\textbf{H}\) learned by the History Encoder which contains rich historical behavior information to compute the type-specific intensity function. Building upon previous explorations of the intensity function~\cite{lin2021empirical,zuo2020transformer}, and taking into account time efficiency as well as the complexity of user behaviors, we adopted a linear interpolation time module to ensure the continuous of the conditional intensity function.  Additionally, we utilized a type-specific nonlinear transformation that transforms the hidden states \(\mathbf{h}(t_{n})\) into a scalar, enhancing the model's capability to capture intricate behavior patterns.
Furthermore, considering the relevance feature of the open-app motivation in \textsection~\ref{sec:data_analysis}, we add the query (recommendation) ratio aware score \(r_{t,m}\). This score captures the proportion of user search or interaction with the recommendation at the current moment \(t\) in the current session. Given that the search ratio is typically low and might introduce noise, we performed batch normalization on \(r_{t,m}\). This normalization step ensures a consistent scale for the ratio, reducing the influence of outliers and allowing for more stable learning.
Finally, we express the intensity function as follows:
\begin{equation}
\lambda_m\left(t \mid \mathcal{T}_t\right) = f\left( \alpha_m \frac{t-t_n}{t_n} + \phi\left( \mathbf{w}^{\top}\mathbf{h}\left(t_n\right) \right) + r_{t,m} \right),
\label{eq:intensity}
\end{equation}
where \(f\) is the softplus function to constrain the intensity function to be positive, \(\alpha_{m}\) is the hyperparameter which modulates the importance of the interpolation, \(\phi\) is the activation function which can be GELU~\cite{hendrycks2016gaussian} or Tanh~\cite{karlik2011performance}, and \(\mathbf{w} \in \mathbb{R}^{d\times1}\) are learnable parameters. For brevity, we omitted batch normalization in~\autoref{eq:intensity}.


\subsection{\mbox{\mymodel: Prediction Layer}}
\label{sec:Open-App Motivation Prediction Layer}
In this section, we introduce the open-app motivation prediction layer to predict the user's next open-app motivation. Drawing inspiration from~\cite{zuo2020transformer}, we found that adding an additional prediction layer on top of the neural Hawkes processes model yields better performance. The prediction layer can integrate user historical information, which contains the short-term user intent \(\textbf{v}_{N}  \) and long-term user intent \(\mathbf{h}(t_{N})\), the time information \(t_{N+1}\) as well as user-specific information \(u\). 

Specifically, we use a time-gate approach to integrate \(\textbf{v}_{N}\) and \(\mathbf{h}(t_{N})\). Inspired by the  relative position encoding methods~\cite{su2021roformer, chen2022time, dai2019transformer}, we use the difference between the current time $t_{N+1}$ and the user's last interaction time \(t_{N,|\mathcal{S}_N |}\) to obtain the relative time \(\Delta t\):
\begin{equation*}
\Delta t = t_{N+1}-t_{N,|\mathcal{S}_N |}.
\end{equation*}
Then, we use the \hyperref[sec:embedding_layer]{\textit{Embedding Layer (6)}} to get the time encoding \(\mathbf{e}_{\Delta t} \in \mathbb{R}^{d}\).
The gating vector can be computed by:
\begin{equation*}
    \mathbf{g}=\operatorname{Sigmoid}\left(\mathbf{W}_l \mathbf{h}(t_{N})+\mathbf{W}_s \textbf{v}_{N}+\mathbf{e}_{\Delta t} \right),
\end{equation*}
where \(\mathbf{W}_l\) and \(\mathbf{W}_s \in \mathbb{R}^{d \times d}\) are the learnable weight matrices.  The integrated history embedding \(\mathbf{z}\) is calculated by:
\begin{equation}
\label{eq:z}
    \mathbf{z} = \mathbf{g} \otimes \textbf{v}_{N} +(1-\mathbf{g}) \otimes \mathbf{h}(t_{N}),
\end{equation}
where \(\otimes\) represents element-wise product. 

To introduce user-specific information, we use \hyperref[sec:embedding_layer]{\textit{Embedding Layer (1)}} to encode the user ID $u$, obtaining \(\mathbf{u}\), which along with \(\mathbf{z}\) is fed into the widely adopted two-layer MLP~\cite{zhou2018deep,zhou2019deep} to model feature interaction. Considering that the sigmoid function is monotonically increasing, we directly add \(r_{N}\) to make use of the relevance feature in open-app motivation:
\begin{equation}
\hat{m}_{N+1}=\operatorname{Sigmoid}\left(\operatorname{MLP}\left(\mathbf{z}\|\mathbf{u}\right)+r_{N}\right),
\end{equation}
where $\|$ denotes the concatenation operation and \(\hat{m}_{N+1}\) denotes the prediction score of \(u\)'s next open-app motivation.

\subsection{Training}
In this section, we describe how our model is trained. For a given user \( u \), in accordance with the existing methods for neural Hawkes process training~\cite{lin2021empirical,zuo2020transformer,mei2017neural}, we maximize the log-likelihood across all the user's history of open-app motivations:
\begin{equation}
\mathcal{L}_{\lambda}^{u}=\frac{1}{N}\sum_{n=1}^{N} \log \lambda_{m_{n}}\left(t_{n} \mid \mathcal{T}_{t_{n}}\right)-\int_{t_1}^{t_{N}} \lambda\left(t \mid \mathcal{T}_t\right) dt
\label{eq:L_lambda}
\end{equation}
where the first term of  equation~\eqref{eq:L_lambda} represents the open-app motivation that happened at the times they happened and the second term is an integral of the total intensities over the observation interval \([t_{1}, t_{N}]\), which can be approximated using Monte Carlo integration~\cite{robert1999monte}. 

For the training of the open-app motivation prediction Layer, we utilize the Binary Cross Entropy loss for optimization:
\begin{equation}
\mathcal{L}_{m}^{u}   = -\frac{1}{N}\sum_{n=1}^{N}(m_{n}   \log(\hat{m}_{n}  ) + (1-m_{n}  ) \log(1-\hat{m}_{n}  )).
\label{eq:pre}
\end{equation}

Finally, for all users in \( \mathcal{U} \), the loss \( \mathcal{L} \) is computed as:
\begin{equation}
\mathcal{L} = \sum_{u\in\mathcal{U}}(-\alpha\mathcal{L}  _{\lambda}^{u}+\mathcal{L}_{m}^{u}   ),
\label{eq:loss}
\end{equation}
where \( \alpha \) are hyperparameters that control the importance of the two parts of loss.

\section{Experiments}
In this section, we empirically verify the efficiency of \mymodel by addressing the following research questions:
\noindent\textbf{RQ1:} How does \mymodel perform in comparison with the baseline models?
\noindent\textbf{RQ2:} Which components of \mymodel are most critical for its performance?
\noindent\textbf{RQ3:} How can \mymodel enhance the performance of downstream tasks?
\noindent\textbf{RQ4:} How do the important hyperparameters in \mymodel affect the model performance?

The source code, \mydata and extended ZhihuRec datasets have been shared at \textcolor{magenta}{\url{https://github.com/Jeryi-Sun/NHP_OAM}}.

\subsection{Experimental Settings}
\subsubsection{Datasets}
\label{sec:dataset}
\begin{table}[t]
\centering
\caption{Statistics of \mydata and ZhihuRec. \# denotes the number of per user; R-C denotes recommended items that were clicked; Q denotes query actions; Q-C denotes clicked items from the search results (ZhihuRec misses this feature); OAM denotes Open-App Motivation. }
\label{tab:data_analysis} 
\begin{tabular}{lcc}
\toprule
\textbf{Information} & \textbf{\mydata} & \textbf{ZhihuRec}\\
\midrule
Users & 25,355 & 798, 086\\
\# R-C, Q, Q-C & 285.73, 17.51, 22.64 & 33.81, 4.89, ---\\
\# Sessions & 20.31 &  5.27 \\
Time span (days) & 16 & 10 \\
\midrule
OAM: search & 66,546 & 715,792 \\
OAM: recommendation & 340,842& 3,314,396\\
\bottomrule
\end{tabular}
\end{table}

\mymodel requires both user session-level S\&R behavior logs and open-app motivations simultaneously. In the following experiments, we evaluated the models on the new real-world Open-App Motivation dataset (\mydata in \textsection~\ref{sec:oamdata describe}) and the extended public S\&R dataset--ZhihuRec~\cite{hao2021large}. As \mydata has already been detailed in Section~\ref{sec:oamdata describe}, this section mainly focuses on how to extend the public S\&R datasets--ZhihuRec. Table~\ref{tab:data_analysis} reports the statistics of both datasets. 

To the best of our knowledge, there doesn't exist a public dataset that contains both explicitly session-level S\&R behaviors and open-App motivation. We extend a public S\&R dataset, the ZhihuRec dataset~\cite{hao2021large}, by generating session-level information and open-app motivations. 
Specifically, consider that ZhihuRec was collected from a fixed day and includes a fixed number of user interactions, while our task requires a continuous history of user interactions. Therefore, we first preprocess ZhihuRec to make it suitable for our task requirements~\footnote{We remove the last day to ensure that users' data distribution is not concentrated on the final day. At the same time, if we don't remove the last day, a large number of users' behaviors would be concentrated in the test set according to our time-based validation and testing method, leading to severe distribution drift.}. Subsequently, we separate a user's entire behavior sequence into sessions, using 30 minutes of inactivity as the interval~\cite{ge2018personalizing,yao2021user}. Finally, we get users' open-app motivations based on whether the user actively searches within 30 seconds of opening the App. If true, the open-app motivation is ``search''; otherwise, it is ``recommendation''.

To more closely approximate real-world application scenarios, we adopt the temporal ordering time-ratio-based splitting strategy~\cite{zhao2022revisiting} to split both datasets. Specifically, for both datasets, to ensure that each user has sufficient history for building a user profile, we treat the log data from the first three days as the historical set. The last day serves as the test set, the second-to-last day as the validation set, and the remaining data as the training set.

\subsubsection{Evaluation Metrics.}
In our specific use case, within Apps such as video platforms that incorporate both Search and Recommendation (S\&R) services, the emphasis is often placed more heavily on the recommendation module. Failing to accurately predict a user's primary motivation for opening the app as \textit{search} can detrimentally impact the user experience. As such, achieving high \textit{precision} in prediction becomes paramount. Consequently, the chosen evaluation metrics include \textbf{Accuracy}, \textbf{Precision}, \( \mathbf{F1}\)\textbf{-Score}, \( \mathbf{F_{0.5}}\)\textbf{-Score}, and \textbf{AUC}~\footnote{ Specifically, \( F_{0.5}\text{-score} \) is a variant of \( F1\text{-score} \) that assigns greater weight to \textit{precision} over \textit{recall}~\cite{ellis2019automatic}.}.

\subsubsection{Baseline Models.}
\label{app:Baseline Models}

Considering the absence of pre-existing models tailored for open-app motivation prediction, we adapt five categories of baseline models to this task: Sequential Recommendation Models, Joint Search-Recommendation Sequential Models, Time-Aware Sequential Recommendation Models, Time-Series Models, and Neural Hawkes Process Models. Each category is modified to predict open motivations, mapping outputs to a [0, 1] interval:

\textbf{Sequential Recommendation Models (SR)}: \textbf{GRU4Rec}~\cite{hidasi2015session} employs Gated Recurrent Units to encode users’ historical sequences.  In our task, we use the same architecture but modify the output layer to a linear layer mapping to [0, 1], with the training objective being
open motivation prediction. \textbf{SASRec}~\cite{kang2018self} employs transformer decoders, adapted similarly. \textbf{BERT4Rec}~\cite{sun2019bert4rec} utilizes transformer encoders, adapted with mask token embedding.

\textbf{Joint Search-Recommendation Sequential Models (SRS)}: \textbf{NRHUB}~\cite{wu2019neural} combines search and recommendation histories with self-attention, adapted with MLP output. \textbf{Query-SeqRec}~\cite{he2022query} merges histories using a transformer encoder. \textbf{USER}~\cite{yao2021user} operates at the session level, combining search and
recommendation history into a single sequence and using
a transformer encoder for encoding.

\textbf{Time-Aware Sequential Recommendation Models (TSR)}: \textbf{TiSASRec}~\cite{li2020time} extends SASRec with interaction timestamps. \textbf{RESETBERT4Rec}~\cite{zhao2022resetbert4rec} enhances BERT4Rec with time information. \textbf{TLSRec}~\cite{chen2022time} is session-based, uses time lag gate, adapted with MLP.

\textbf{Time-Series Models (TS)}: \textbf{LSTNet}~\cite{lai2018modeling} captures dependencies in time-series data with CNN and RNN layers. \textbf{Autoformer}~\cite{chen2021autoformer} Uses series decomposition to obtain trend and seasonal information, relying on FFT and IFFT-based Auto-correlation. \textbf{DLinear}~\cite{zeng2023transformers} decomposes time series, predicts with FFN.

\textbf{Neural Hawkes Process Models (NHP)}: \textbf{CTNHP}~\cite{mei2017neural} models events in continuous time with LSTM. \textbf{SAHP}~\cite{zhang2020self} applies self-attention for event prediction. \textbf{THP}~\cite{zuo2020transformer} employs self-attention in point-process-based RNN models.

    
    
    
    
    

\begin{table*}[t]
\caption {Performance comparisons between \mymodel  and the baselines. The boldface represents the best performance. `$\dagger$' indicates that the improvements over all of the baselines are statistically significant (t-tests, $p\textrm{-value}< 0.05$). 
}\label{tab:main results} 
\centering
\resizebox{0.99\linewidth}{!}{
\begin{tabular}{llcccccccccc}
\toprule 

\multirow{2}{*}{\textbf{Types}}&\multirow{2}{*}{\textbf{Models}} & \multicolumn{5}{c}{\textbf{\mydata}} & \multicolumn{5}{c}{\textbf{ZhihuRec}}                                                                      \\
\cmidrule(lr){3-7}\cmidrule(lr){8-12}
  & &\textbf{Accuracy} & \textbf{Precision}  & \textbf{F1-Score} & $\textbf{F}_{0.5}$\textbf{-Score} & \textbf{AUC-Score} & \textbf{Accuracy} & \textbf{Precision}  & \textbf{F1-Score} & $\textbf{F}_{0.5}$\textbf{-Score} & \textbf{AUC-Score} \\
\cdashline{1-12}
\multirow{3}{*}{\textbf{SR}} & SASRec & 0.9193&	0.8156&	0.4941&	0.6472	& 0.7162& 0.7247 & 0.5431  & 0.5352 & 0.5399 & 0.6611  \\ 
        &BERT4Rec & 0.9170	&0.8310	&0.4602	&0.6285	&0.7269& 0.7551 & 0.6145  & \underline{0.5491} & 0.5865 & 0.6826  \\ 
        &GRU4Rec & 0.9228 & 0.8789	& 0.5046 &0.6778	&0.7181 & 0.7521 & 0.6061  & 0.5479 & 0.5814 & 0.6827  \\ \cdashline{1-12}
        \multirow{3}{*}{\textbf{SRS}}&NRHUB & 0.9055	&0.7289	&0.3384	&0.4987&	0.7336 & 0.7451&0.6023&0.5294&0.5709&0.6853\\ 
        &Query-SeqRec & 0.9166	&0.7626	&0.4783	&0.6161	&0.8012 & 0.7532 & 0.6120  & 0.5429 & 0.5823 & 0.7237  \\ 
        &USER & 0.9103	&0.7304&	0.4136&	0.5591	& 0.7986 & 0.7533 & 0.6132  & 0.5410 & 0.5821 & \underline{0.7254}  \\ \cdashline{1-12}
        \multirow{3}{*}{\textbf{TSR}}&TiSASRec & 0.9164	&0.7387&	0.4908&	0.6145&	0.8100& 0.7501 &	0.6007 & 0.5468	&0.5779	& 0.6919 \\ 
        &RESETBERT4Rec & 0.9159	&0.7429	&0.4824&0.6109	&0.8117& \underline{0.7552} & \underline{0.6146}  &\underline{0.5491} & \underline{0.5866} & 0.6937  \\ 
        &TLSRec & 0.9107&	0.7119	&0.4334&	0.5664&	0.7982& 0.7547 & 0.6131  & \textbf{0.5494} & 0.5859 & 0.6997  \\ \cdashline{1-12}
        \multirow{3}{*}{\textbf{TS}}&Dlinear & 0.9221&	0.7743	&0.5467	&0.6638	&0.8316 & 0.6182 & 0.3940  & 0.4419 & 0.4118 & 0.5871  \\ 
        &Autoformer & 0.9120	&0.6502	&0.5322	&0.5972	&0.8171 & 0.6518 & 0.4439  & 0.5201 & 0.4715 & 0.6757  \\ 
        &LSTNet & 0.9226	&0.7973	&0.5385	&0.6688	&0.8352& 0.7116 & 0.5190  & 0.5338 & 0.5248 & 0.7003  \\ \cdashline{1-12}
        \multirow{3}{*}{\textbf{NHP}}&THP  & 0.9393&	\underline{0.9019}&	0.6669&	\underline{0.8050}&	0.9063& 0.7163 & 0.5273 & 0.5321 & 0.5292 & 0.7064  \\ 
        &CTNHP & \underline{0.9409} & 0.8883	&\underline{0.6912}	&0.7974	&0.9017 & 0.7161 & 0.5268 &  0.5336 & 0.5295 & 0.7085  \\ 
        &SAHP & \underline{0.9409}&	0.9014&	0.6638&	0.7885&	\underline{0.9076} & 0.7153 & 0.5253 & 0.5343 & 0.5289 & 0.7064  \\ \cdashline{1-12}
        \textbf{Ours} &\mymodel & \textbf{0.9463}$^\dagger$	& \textbf{0.9320}$^\dagger$&	\textbf{0.7100}$^\dagger$&	\textbf{0.8284}$^\dagger$	&\textbf{0.9326}$^\dagger$ & \textbf{0.7601}$^\dagger$& \textbf{0.6326}$^\dagger$  & 0.5462 & \textbf{0.5950}$^\dagger$ & \textbf{0.7541}$^\dagger$ \\ 
\bottomrule
\end{tabular}
}
\end{table*}

\subsubsection{Implementation details}
\label{app:implementation_details}
\mymodel's hyperparameters are tuned using grid search on the validation set with Adam~\cite{kingma2014adam}. 
The batch size is tuned among $\{64, 128, 256\}$. The learning rate $\eta$ is tuned among $\{1e-5, 1e-4, 1e-3\}$. The embedding dimension $d$ is set to 32. The number of history encoder layers  $L_{1}$ and $L_{2}$ are both set to 2.   The $\alpha$ is tuned among $\{0.1, 0.001, 0.0001, 0.00001, 0.000001\}$. For all models, the maximum session number is set to 10 on the \mydata dataset and 5 on the ZhihuRec dataset. The maximum sequence length of behaviours in a session is set to 20 on the \mydata dataset and 5 on the ZhihuRec dataset.
Since the output of  \mymodel falls within the range \( (0, 1) \), we need to establish a threshold within $(0, 1)$  for classification purposes. In our open-app motivation prediction task, we opt for the \( F_{0.5} \)-score as our evaluation metric because we place a greater emphasis on precision while also considering recall. The threshold is determined based on achieving the optimal \( F_{0.5} \)-score on the validation set. The search range for the threshold starts from 0 and increases by increments of 0.01, up to and including 1. The optimal threshold is then selected from these candidates, which achieves the highest \( F_{0.5} \)-score on the validation set, to be applied to the test set for final classification results.
For baselines, we tuned the parameters using the grid search around the optimal values in the original paper.

\subsection{Overall Performance (RQ1).}

Table~\ref{tab:main results} shows prediction performances of \mymodel versus baselines on \mydata and ZhihuRec. Based on the results presented in~\autoref{tab:main results}, we found:
    (1) \mymodel almost outperformed the baseline models in terms of five evaluation metrics on \mydata and ZhihuRec, with statistical significance (t-tests, \(p\)-value < 0.05). These results verified the efficiency of \mymodel in predicting users' open-app motivations.
    (2) \mymodel outperformed Time-Aware Sequence Recommendation Models and Sequence Recommendation Models, demonstrating that relying solely on history-level information was insufficient. It was also essential to model and fully utilize the behavioral patterns behind users' open-app motivations. At the same time, Time-Aware Sequence Recommendation Models and Sequence Recommendation Models outperformed Sequence Recommendation Models, indicating the importance of incorporating time information and unified modeling for search and recommendation in predicting open-app motivations.
    (3) \mymodel and Neural Hawkes Process Models outperformed Time-Series Models, proving that the neural Hawkes process effectively captured the features found in past open-app motivations without needing to assume consistent sampling time intervals like time-series models.
    (4) \mymodel performed better than the Neural Hawkes Process baselines, showing the effectiveness of our improvements in applying the Hawkes Process to open-app motivation prediction.
    (5) The results on ZhihuRec are generally lower than those on OAMD, mainly because ZhihuRec is adapted to perform this task.

\begin{table}[t]
\caption{Ablation Study of \mymodel on \mydata.}
\label{tab:ablation_table}
\centering
\resizebox{0.99\linewidth}{!}{
    \begin{tabular}{lccccc}
    \toprule 
    \textbf{Models} & \textbf{Accuracy} & \textbf{Precision}  & \textbf{F1-Score} & $\textbf{F}_{0.5}$\textbf{-Score} & \textbf{AUC-Score} \\
    \cdashline{1-6} 
        w/o SLE  & 0.9436 & 0.9003 & 0.6858 & 0.8002 & 0.9144  \\ 
        w/o UE & 0.9460 & 0.9123 & 0.7077 & 0.8177 & 0.9323  \\ 
        w/o TG & 0.9448	& 0.8965 & 0.6960 &	0.8039 & 0.9307\\
        w/o PL & 0.9433	& 0.9056 & 0.6817&	0.8005	& 0.9289\\
        w/o QR & 0.9457 & 0.9153 & 0.6975 & 0.8137 & 0.9302  \\ 
        w/o HP & 0.9372 & 0.8672 & 0.6452 & 0.7623 & 0.9212 \\ 
        \cdashline{1-6} 
        \mymodel & \textbf{0.9463} & \textbf{0.9320} & \textbf{0.7100} & \textbf{0.8284} & \textbf{0.9326}  \\      
    \bottomrule
    \end{tabular}
}
\end{table}
\subsection{Ablation Study (RQ2)}

\mymodel consists of several key components, and to understand the effects of each component, we conducted several ablation experiments for \mymodel.  The studies involved removing the Session-level Encoder (\textbf{w/o SLE}), excluding the query (recommendation)  ratio aware score (\textbf{w/o QR}), using LSTM in place of the Hawkes process (\textbf{w/o HP}), removing user-specific information (\textbf{w/o UE}), eliminating the time-gate (\textbf{w/o TG}), and removing the Prediction Layer (\textbf{w/o PL}). 

Table~\ref{tab:ablation_table} presents the performance of various model variants on the \mydata~\footnote{
To ensure the accuracy of the experiment, we conducted a series of analytical experiments on the real-world dataset \mydata.}. The results reveal several insights: First, the performance of the model without the Session-level Encoder (w/o SLE) falls short of that of \mymodel, emphasizing the importance of capturing rich contextual information within a user's session for improved model effectiveness. Second, the absence of the query (recommendation) ratio aware score (w/o QR) also leads to inferior performance, which underscores the significance of modeling relevance features. Third, excluding the Hawkes process (w/o HP) results in a significant drop in performance, highlighting the advantages of using this process to model open-app motivation. Lastly, we observe a decline in performance when User-specific information (w/o UE), Time-gate (w/o TG), and Prediction Layer (w/o PL) are removed, demonstrating the efficacy of using the Prediction Layer to model both time information and user-specific information.



\subsection{Application Experiments (RQ3)}
In this section, we evaluate the performance of two downstream recommendation tasks when deploying \mymodel with them. Specifically, the two applications are Item Recommendation and Open-App Item Recommendation. The performance of \mymodel on these applications reflects its effectiveness in learning better user representations, especially for tasks that require arousing user interest in a very short time,  highlighting the immense application value of \mymodel.

\subsubsection{Item Recommendation}
\begin{figure}
    \centering
    \begin{subfigure}{0.49\linewidth}
        \centering
        \includegraphics[width=\textwidth]{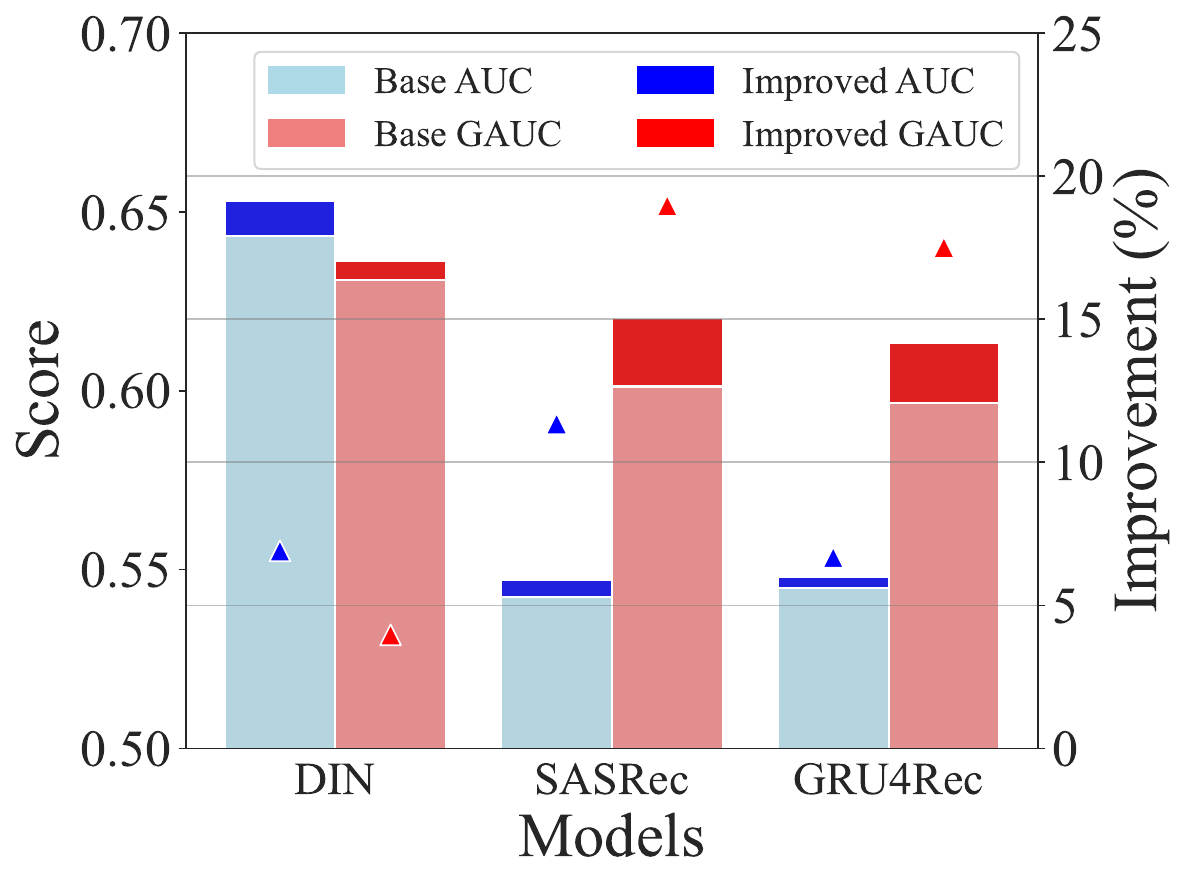}
        \caption{SR Models}
        \label{subfig:SR_app_1}
    \end{subfigure}
    \hfill
    \begin{subfigure}{0.49\linewidth}
        \centering
        \includegraphics[width=\textwidth]{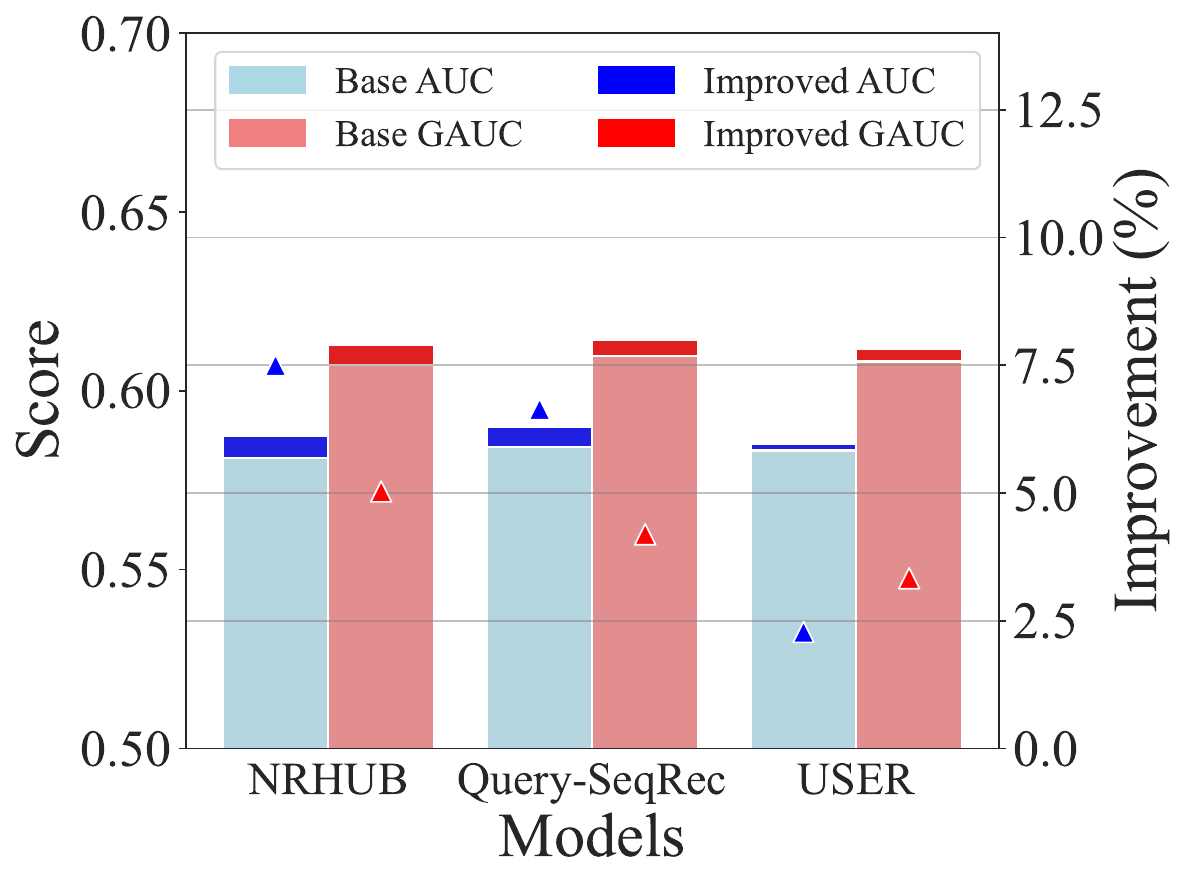}
        \caption{SRS Models}
        \label{subfig:SRS_app_1}
    \end{subfigure}
    
    \caption{Application on Item Recommendation.}
    \label{fig:app_1}
\end{figure}


Item Recommendation aims to generate a list of recommendations for a user \( u \). We utilized the embedding \( \mathbf{z  } \) in~\autoref{eq:z} trained by \mymodel to improve the item recommendation models~\cite{zhou2018deep, kang2018self, hidasi2015session}, which can be formulation as:
\begin{equation*}
    p(i | u) = \sigma(\boldsymbol{i}^{\top} \cdot \boldsymbol{u}),
\end{equation*}
where \( \boldsymbol{u} = \text{FFN}([\mathbf{u}_{reco}; \mathbf{z  }]) \), \( [\cdot ; \cdot] \) concatenates vectors and $\sigma$ is the Sigmoid function. The \( \textbf{u}_{reco} \) vector is sourced from the original Item Recommendation models. For Joint Search-Recommendation Sequential Models~\cite{wu2019neural,he2022query,yao2021user}, we can also utilize the predicted open-app motivation score to control the weight of Search items and Recommendation items within the current session history. For the dataset, we use items clicked by users in \mydata as positive items and unclicked items as negative items, adopting the same splitting method as in \textsection~\ref{sec:dataset}, which aligns well with practical applications. Evaluation metrics include AUC, GAUC, and we separately calculate the relative improvement after applying \mymodel followed~\cite{zhou2018deep}. The item recommendation model and \mymodel are trained simultaneously. Evidenced by Figure~\ref{fig:app_1}, there is a noticeable improvement, serving as an indicator of our model's effectiveness in learning better user representations.

\subsubsection{Open-App Item Recommendation}


\begin{figure}
    \centering
    
    \begin{subfigure}{0.49\linewidth}
        \centering
        \includegraphics[width=\textwidth]{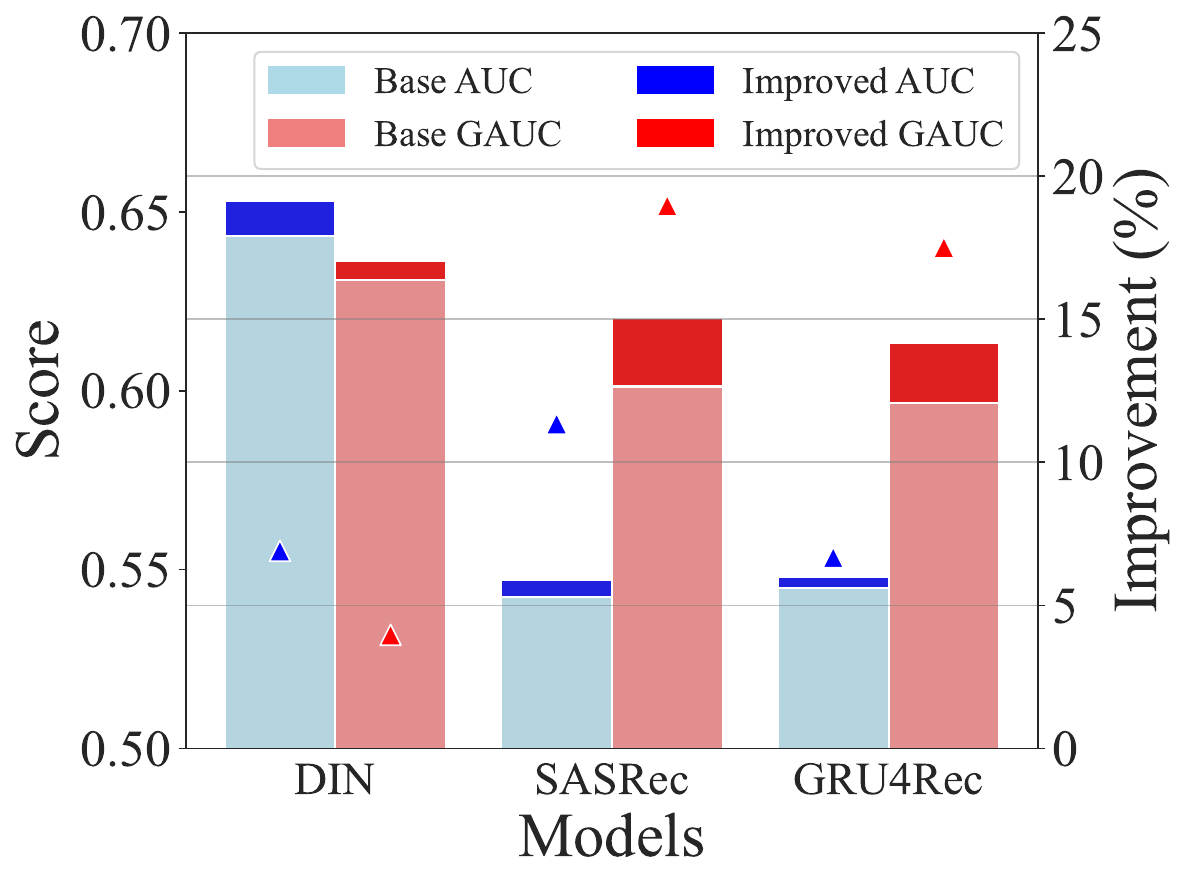}
        \caption{SR Models}
        \label{subfig:SR_app_2}
    \end{subfigure}
    \hfill
    \begin{subfigure}{0.49\linewidth}
        \centering
        \includegraphics[width=\textwidth]{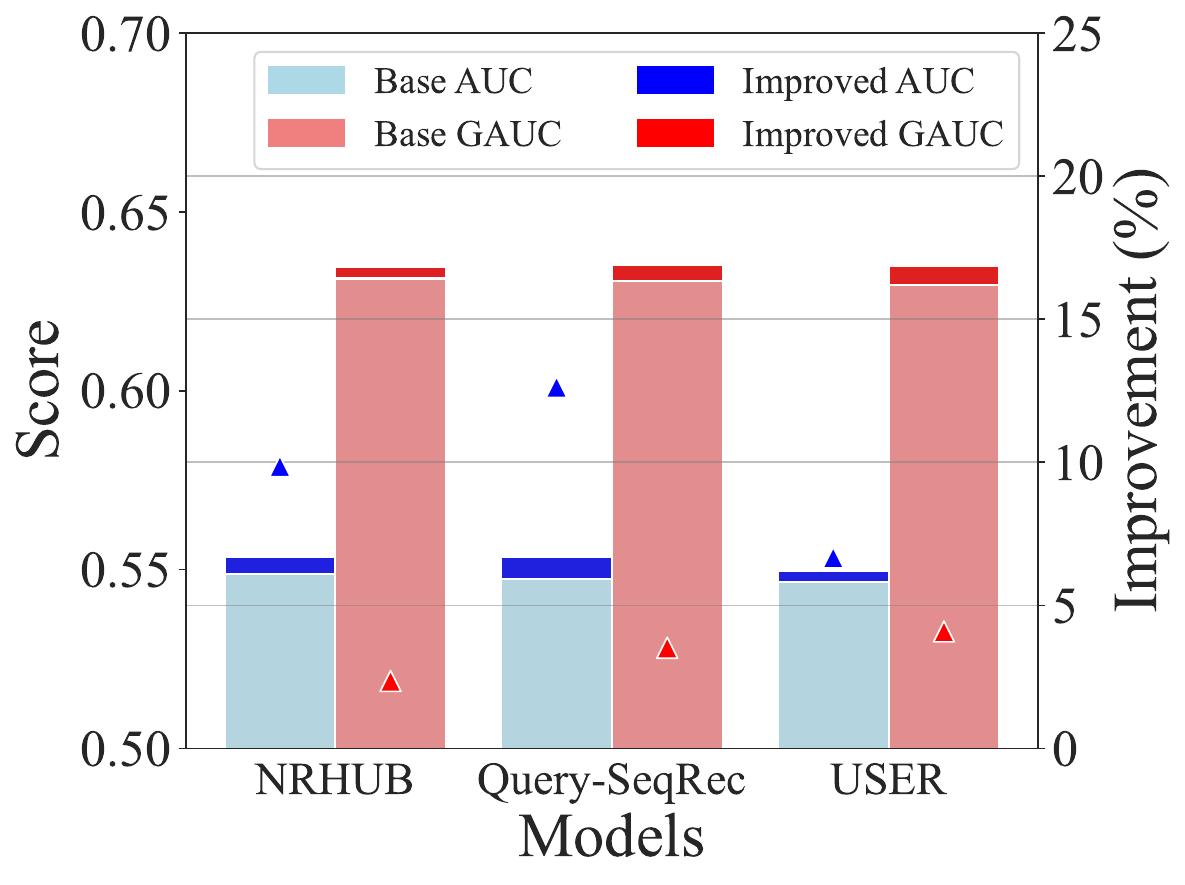}
        \caption{SRS Models}
        \label{subfig:SRS_app_2}
    \end{subfigure}
    \caption{Application on Open-App Item Recommendation.}
    \label{fig:app_2}
\end{figure}

Open-App Item Recommendation refers to a type of Item Recommendation that has only a single opportunity to recommend, such as App Open Advertisement Recommendation and Message Pop-up Recommendation~\cite{lan2023neon}. We employed the same model architecture, evaluation metrics, and training set as used in the above Item Recommendation. For the test set, we selected items from the user's first interaction, which includes both searched and recommended items. An interaction is considered a positive item if it is clicked, and a negative item otherwise. As evident from Figure~\ref{fig:app_2}, there is a significant improvement, which serves as evidence for \mymodel's effectiveness in capturing users' immediate interests upon app opening.

\subsection{Impact of Session Length (RQ4)}
\begin{figure}
    \centering
    \includegraphics[width=0.95\linewidth]{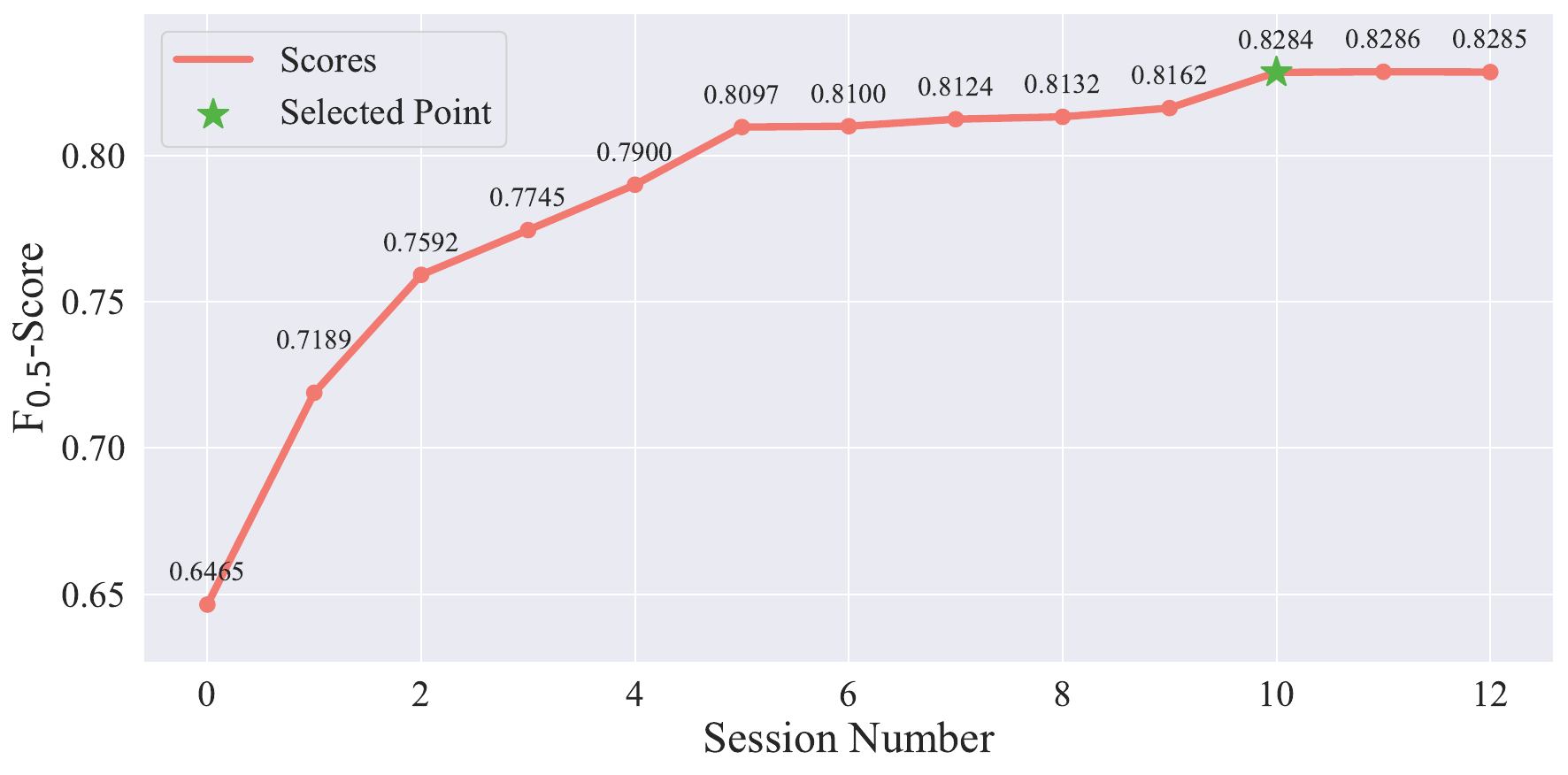}
    \caption{Performance by Session Numbers on \mydata.}
    \label{fig:performance_session_len}
\end{figure}

In light of the Hawkes process's sensitivity to sequence length, we systematically explored its impact on the \( F_{0.5}\text{-Score} \) in \mymodel on \mydata. Our experimental setup covered a range of session numbers from 0 to 12. As illustrated in Figure~\ref{fig:performance_session_len}, we observed a significant increase in \( F_{0.5}\text{-Score} \) as the session number increased, stabilizing at a session number of 10. This result justifies our choice of 10 as the optimal session number for our model. This stabilization behavior is consistent with the known characteristics of the Hawkes process, which often requires adequate sequence lengths to accurately model complex systems. Hence, our findings validate our model's alignment with the behavior expected from the Hawkes process.


\section{CONCLUSIONS}
In this work, we introduce a novel problem, \emph{open-app motivation prediction}, which is pivotal for platforms providing both search and recommendation services. We propose the \mymodel model, effectively tackling associated challenges. Leveraging the Neural Hawkes Process (NHP), our model captures temporal patterns and employs a hierarchical Transformer architecture, along with an intensity function aware of the relevance feature. We also construct a new real-world dataset, called \mydata, to advance the study of open-app motivation prediction. Comprehensive experiments demonstrate significant improvement over baseline models, underscoring the considerable application value of \mymodel.


\bibliographystyle{ACM-Reference-Format}
\bibliography{sample-base}
\newpage
\appendix

\end{document}